\documentstyle[preprint,prl,aps]{revtex}
\begin{document}
\draft
%
%
\title{Against ``Against Many-Worlds Interpretations''}
%
%
\author{Toshifumi Sakaguchi}
%
%
\address{
ASCII Corporation \\
Yoyogi 4-33-10, Shibuya-Ku, Tokyo, Japan
}
\date{\today}
\maketitle
%
%
\begin{abstract}
The paper entitled ``Against Many-Worlds Interpretations'' by A.~Kent,
which has recently been submitted to the e-Print archive (gr-qc/9703089)
contained some misconceptions.  The claims on Everett's many-worlds
interpretation are quoted and answered.
\end{abstract}
%
%
\narrowtext
%
%
A.~Kent has submitted the paper~\cite{rf:kent} entitled ``Against
Many-Worlds Interpretations'', in which we can see frequently (almost
endlessly) raised claims on Everett's many-worlds interpretation (MWI).
However, these claims apparently came from misconceptions about
Everett's original MWI and I think we should correct them.  The paper
neither is intended for a debate on the different versions of MWIs, nor
criticize recent post-Everettians view, such as consistent histories
approach.  I just show what the Everett MWI is.

In the paper~\cite{rf:kent} (Subsection ``A.~Everett'' in Section II.
``THE CASE AGAINST''), the author tries to illustrate problems on the
Everett MWI~\cite{rf:everett} by a simple example (typographic errors
are corrected, all quotes appear in italics):
\begin{quote}
{\it
Suppose that a previously polarized spin-$\frac{1}{2}$ particle has just
had its spin measured by a macroscopic Stern-Gerlach device, on an axis
chosen so that the probability of measuring spin $+\frac{1}{2}$ is
$\frac{2}{3}$. The result can be idealized by the wavefunction
\begin{equation}
\phi = a \, \phi_0 \otimes \Phi_0 + b \, \phi_1 \otimes \Phi_1
\eqnum{2}
\label{eq:2}
\end{equation}
where $\phi_0$ is the spin $+\frac{1}{2}$ state of the particle,
$\Phi_0$ the state of the device having measured spin $+\frac{1}{2}$;
$\phi_1$, $\Phi_1$ likewise correspond to spin $-\frac{1}{2}$;
$|a|^2 = \frac{2}{3}$, $|b|^2 = \frac{1}{3}$.  Now in trying to interpret
this result we encounter the following problems: $\ldots$
}
\end{quote}
First of all, the above wavefunction does not describe any {\em
quantum-mechanical measurement} process which is distinguished from the
classical one by the occurrence of the collapse-of-wavefunction (COW)
{\em phenomenon}.  It simply describes a state after the interaction
between the spin and the Stern-Gerlach device occurred.  Quantum
mechanics cannot contradict with the above description, even the
Copenhagen interpretation agrees with until an observation by an
observer takes place~\cite{rf:neumann}.

In Everett's MWI, it is essential to consider an {\em observer} state,
in which all physical informations which were observed by the observer
are embedded.  Therefore, the above equation should be replaced by
\begin{equation}
\phi = a \, \phi_0 \otimes \Phi_0 \otimes \Psi_0
 + b \, \phi_1 \otimes \Phi_1 \otimes \Psi_1
\eqnum{2a}
\label{eq:2a}
\end{equation}
where, $\Psi_0$ and $\Psi_1$ denote the observer states who observed the
spin $+\frac{1}{2}$ and $-\frac{1}{2}$ respectively.  This is a direct
consequence of (A) the linearity of the time evolution operator $U(t)$
which includes the interaction between the object and observer systems
and of (B) the relation which should be satisfied if the object system
is prepared in the {\em eigenstates} $\phi_i$ and the observer system is
prepared to observe the observable (the spin component):
\begin{equation}
U(t) \phi_i \otimes \Phi_i \otimes \Psi
= \phi_i \otimes \Phi_i \otimes \Psi_i.
\eqnum{2b}
\label{eq:2b}
\end{equation}

The author continues:
\begin{quote}
{\it
Firstly, no choice of basis has been specified; we could expand $\phi$
in the 1-dimensional basis $\{ \phi \}$ or any of the orthogonal
2-dimensional bases
\begin{equation}
\{ \cos \theta \, \phi_0 \otimes \Phi_0  +
\sin \theta \, \phi_1 \otimes \Phi_1 ,
\sin \theta \,  \phi_0 \otimes \Phi_0   -
\cos \theta \, \phi_1 \otimes \Phi_1 \}
\eqnum{3}
\label{eq:3}
\end{equation}
or indeed in multi-dimensional or unorthogonal bases. Of course, the
information is in the wavefunction is basis-independent, and one is free
to choose any particular basis to work with.  But if one intends to make
a physical interpretation only in one particular basis, using quantities
(such as $|a|^2$ and $|b|^2$) which are defined by that basis, one needs
to define this process (and, in particular, the preferred basis) by an
axiom. This Everett fails to do.
}
\end{quote}
The author apparently fails in understanding what the Everett MWI is.
As stated above, Eq.\ (\ref{eq:2}) by itself tells nothing about what
was observed by an observer unless we specify the observer.

In the state (\ref{eq:2a}), we can see that there are two observer
states $\Psi_0$ and $\Psi_1$, each of which is evolved from the same
observer state $\Psi$ as shown in Eq.\ (\ref{eq:2b}).  These two
observer states correspond to the two observers seeing different
outcomes, i.e., spin+$\frac{1}{2}$ and spin-$\frac{1}{2}$.  There is no
{\em objective} COW, but it {\em appears} to the observer that the state
of the {\em object} system has collapsed into one of the eigenstates of
the observable, while the whole system which includes the observer
system is still in a superposition.  This is the Everett MWI.

We do not have to require any loss of coherence between the branched
states: The process described by Eq.\ (\ref{eq:2b}), in which there is
no room for indicating any kind of decoherence, defines the observer
states unambiguously.  This is why there is no need for a preferred
basis.  (Given a Hamiltonian, eigenstates of the object system, initial
state of the measurement device, and initial state of the observer
system, we can check if Eq.\ (\ref{eq:2b}) holds true.)

The following claim comes from author's misconception.
\begin{quote}
{\it
Secondly, suppose that the basis
$(\phi_0 \otimes \Phi_0 , \phi_1 \otimes \Phi_1 )$ is somehow selected.
Then one can perhaps {\em intuitively} view the corresponding components
of $\phi$ as describing a pair of independent worlds. But this intuitive
interpretation goes beyond what the axioms justify; the axioms say
nothing about the existence of multiple physical worlds corresponding to
wavefunction components.
}
\end{quote}
As explained above, there is no {\em objective} multiple physical worlds
corresponding to wavefunction components. The world is subjective and
relative to each observer. Let us go on to the next claim:
\begin{quote}
{\it
Thirdly, in any case, no physical meaning has been attached to the
constants $|a|^2$ and $|b|^2$.  They are not to be interpreted as the
probabilities that their respective branches are realized; this is the
whole point of Everett's proposal. It can not be said that a proportion
$|a|^2$ of the total number of worlds is in state
$\phi_0 \otimes \Phi_0$; there is nothing in the axioms to justify this
claim. (Note that if the two worlds picture {\em were} justified, then
each state would correspond to one world, and it must be explained why
each measurement does not have probability $\frac{1}{2}$.) Nor can one
argue that the probability of a particular observer finding herself in
the world with state $\phi_0 \otimes \Phi_0$ is $|a|^2$; this conclusion
again is unsupported by the axioms.
}
\end{quote}
The probability is {\em not} a branching rate into many-worlds, but is the
relative frequency which is counted by the observer in {\em each}
world~\cite{rf:sakaguchi1}.  It is easy to show that the observer who is
described in {\em each} element of the {\em superposition} agrees that
the relative frequency equals to $|a|^2 : |b|^2$ in the limit that the
number of trials goes to infinity.  (Quantum mechanics tells nothing
about the outcome of each trial:  No one can see any interference
pattern in a double slit experiment with only a few of dots on the
screen.)

Finally, the author considers the {\em measure} and claims:
\begin{quote}
{\it
$\ldots$
$\mu (\Phi^{\epsilon})=0 $ cannot imply that $\Phi^{\epsilon}$ is
physically irrelevant, when no hypothesis has been made about the
physical meaning of $\mu$.
}
\end{quote}
where, $\mu (\Phi^{\epsilon})$ is introduced by the author as a measure.
However, it is not the Everett measure, but is a norm which is equal to
$(\mbox{Everett measure})^{1/2}$ (see author's claim below Axiom 1E).
The claim seems to arise from a confusion of two totally different
concepts; ``norm of a state vector'' and ``measure of a subset in a
superposition''.

In order to derive {\em quantitative} results in a theory, we need to
have a measure. Even in a classical theory, we have a measure, though it
is trivial one~\cite{rf:everett}.  If we have no measure in a theory
there is no hope of finding any quantitative result in it (think of
measure in path integral, for example). What Everett showed in his paper
is that we can construct a consistent quantum theory of a closed system
{\em if} we take the Everett measure.
%
%

%
%
%
\end{document}